\documentstyle[12pt]{article}
\begin{document}
\tolerance=5000
\def\be{\begin{equation}}
\def\ee{\end{equation}}
\def\bea{\begin{eqnarray}}
\def\eea{\end{eqnarray}}
\def\nn{\nonumber \\}
\def\cF{{\cal F}}
\def\det{{\rm det\,}}
\def\Tr{{\rm Tr\,}}
\def\e{{\rm e}}
\def\etal{{\it et al.}}
\def\erp2{{\rm e}^{2\rho}}
\def\erm2{{\rm e}^{-2\rho}}
\def\er4{{\rm e}^{4\rho}}
\def\etal{{\it et al.}}

\  \hfill
\begin{minipage}{2.5cm}
NDA-FP-61 \\
May 1999
\end{minipage}

\

\vfill

\begin{center}

{\large\bf  Thermodynamics of Schwarzschild-(Anti-)de Sitter 
Black Holes with account of quantum corrections}

\vfill

{\large\sc Shin'ichi NOJIRI}\footnote{
e-mail : nojiri@cc.nda.ac.jp}
and
{\large\sc Sergei D. ODINTSOV$^{\spadesuit}$}\footnote{
e-mail : odintsov@mail.tomsknet.ru, sergei.odintsov@itp.uni-leipzig.de}
\vfill

{\large\sl Department of Mathematics and Physics \\
National Defence Academy \\
Hashirimizu Yokosuka 239, JAPAN}

{\large\sl $\spadesuit$
Tomsk Pedagogical University \\
634041 Tomsk, RUSSIA \\
}

\vfill

{\bf ABSTRACT}

\end{center}

We discuss the quantum corrections to thermodynamics 
(and geometry) of S(A)dS BHs using large $N$ one-loop 
anomaly induced effective action for dilaton coupled matter 
 (scalars and spinors). It is found the temperature, mass 
and entropy with account of quantum effects for 
multiply horizon SdS BH and SAdS BH what also gives the corresponding 
expressions for their limits: Schwarzschild and de Sitter spaces. 
In the last case one can talk about quantum correction to 
entropy of expanding Universe. 

The anomaly induced action under discussion corresponds to 
4d formulation ($s$-wave approximation, 4d quantum matter is minimal 
one) as well as 2d formulation (complete effective action, 2d quantum 
matter is dilaton coupled one). Hence, most of results are given 
for the same gravitational background with interpretation as 
4d quantum corrected BH or 2d quantum corrected dilatonic BH.
Quantum aspects of thermodynamics of 4d 't Hooft BH model
are also considered.

\newpage

\section{Introduction}

The famous Bekenstein-Hawking entropy of black hole
 \cite{BH} is known to be proportional to the surface area 
of its event horizon. It gives the bright manifestation of
analogy between thermodynamics and BHs \cite{BCH}. However, despite 
numerous attempts varying from strings \cite{SV}, 
three dimensional gravity \cite{SC,BZ}, induced gravity 
 \cite{TJ}, etc., the derivation of BH entropy 
from statistical mechanics is not yet completely clear.

Among other (thermodynamic) quantities used to describe BHs 
one can mention not only entropy but also 
temperature, horizon radius, mass, energy, charges,etc.
In order to understand better the thermodynamic properties 
of BHs it is extremely important to find the unified way 
to evaluate the quantum corrections to all above 
characteristics of BH. In its own turn, such study is expected to  
give the insights to the better formulations of quantum gravity.
 
Unfortunately, the calculation of quantum corrections 
to thermodynamics of different BHs is not easy task. 
Usual approach is to consider (simple) quantum matter 
on some BH background (as a rule Rindler space) 
where spectrum of the corresponding operator is known. 
Then, in frames of some regularization one can 
find the effective action (or stress energy tensor) 
what may be used as quantum correction to classical action.
This very tedious procedure should be repeated again 
for each specific geometry (when it is possible). 
The complete effective action (EA) on an arbitrary 
gravitational background is not yet known (see ,for a review 
\cite{5}). It would be really interesting to present 
more universal approach, at least in the situations
 when EA is known for some classes of backgrounds.

In the present paper we suggest such universal 
prescription to the calculation of quantum corrections 
to BHs. It is based on anomaly induced EA for 2d 
dilaton coupled matter (scalars and spinors). Such EA 
is found on an arbitrary dilaton-gravitational 
background. From 4d point of view it gives so-called 
s-wave approximation EA and from 2d point of view 
it gives total one-loop EA for dilaton coupled matter 
which is typical in 2d Brans-Dicke-matter theory 
(in Einstein frame). As dilaton and metric dependence 
of this EA is known it may be easily added to classical action.
After that one is left with modified but still CLASSICAL 
gravity where calculation of thermodynamics of BHs is 
now routine work. Moreover, the results obtained in such way 
have 4d interpretation as well as 2d (4d BH may be easily 
interpreted as 2d dilatonic BH). 

We mainly discuss multiply horizon Schwarzschild-(Anti-)de Sitter 
(S(A)dS) BHs which typically may occur in
the early Universe as primordial ones. In addition, as 
limiting backgrounds they have de Sitter and Schwarzschild 
spaces. Note that S(A)dS BHs do not normally appear
at the final stage of star collapse. Nevertheless, 
still there maybe mechanisms to produce them:
via BHs creation at the early Universe \cite{BSWH} or 
via direct inducing of such primordial objects due to
quantum effects \cite{byts}. 

They also may demonstrate the realization of beautiful 
BHs anti-evaporation process \cite{SWH,Noj} which may 
become the basis for creation of
 muliply inflationary Universes \cite{SWH,byts}. It is 
important that anti-evaporation of BHs may put 
cosmological limits to the content of GUTs\cite{Noj}.

The paper is organised as follows. The review of
anomaly induced  EA evaluation 
and construction of effective equations of motion is presented 
in the next section. In section three, using such EA 
we find quantum corrections to BH entropy, mass, 
horizon radius and temperature for 4d and 2d Schwarzschild BHs. 
In section 4 the same problem is solved for SdS (or SAdS) BHs.
The results of previous section may be reproduced from 
such calculation by taking the correspondent limit. 
Quantum properties of another limit (de Sitter space)
are also discussed.  In this case we get the quantum correction 
to the entropy of expanding Universe. In the last section 
we discuss quantum corrections to thermodynamics of 
't Hooft BH model, working in the same fashion. 
Some outlook related with quite general character 
of our approach as well as the possibility of other applications 
is given in final section.

\section{The effective action and equations of motion}

We will start from the action of Einstein gravity with
$N$ minimal real scalars and $M$ Majorana fermions
\bea
\label{e1}
S_{4d}&=&-\frac{1}{16\pi G}\int
d^4x\,\sqrt{-g_{(4)}} \left(R^{(4)}-2\Lambda\right) \nn
&& + \int d^4x\,\sqrt{-g_{(4)}}\,
\left(\frac{1}{2}\sum_{i=1}^Ng^{\alpha\beta}_{(4)}
\partial_\alpha \chi_i \partial_\beta \chi_i + \sum_{i=1}^M 
\bar\psi_i \gamma^\mu\nabla_\mu \psi_i \right)
\eea
where $\chi_i$ and $\psi_i$ are real scalars and Majorana spinors,
respectively. In order to apply large $N$ approach,  $N$ and $M$ 
are considered to be large, $N,M \gg 1$, $G$
and $\Lambda$ are gravitational and cosmological constants, 
respectively.

The convenient choice for the spherically symmetric spacetime is the
following one
\be
\label{e2}
ds^2=g_{\mu\nu}dx^\mu dx^\nu+e^{-2\phi}d\Omega,
\ee
where  $\mu,\nu=0,1$, $g_{\mu\nu}$ and $\phi$ depend only on  $x^0$,
$x^1$ and $d\Omega$ corresponds to the two-dimensional sphere.

The action (\ref{e1}), reduced according to (\ref{e2}) takes 
the form
\bea
\label{e3}
S_{red}&=&\int d^2x \sqrt{-g}\e^{-2\phi}
\left[-{1 \over 16\pi G}
\{R + 2(\nabla  \phi)^2 -2\Lambda + 2\e^{2\phi}\} \right. \nn
&& \left. + {1 \over 2}\sum_{i=1}^N(\nabla \chi_i)^2
+ \sum_{i=1}^{2M} \bar\psi_i \gamma^\mu \nabla_\mu \psi_i \right]
\eea
Note that the fermion degrees of freedom 
after reduction are twice of original ones. 

Working in large $N$ and $s$-wave approximation, 
one can calculate the quantum correction to $S_{red}$ 
(effective action). Using 2d conformal anomaly for dilaton coupled 
scalar, calculated in \cite{1} (see also
\cite{2,3,4}) one can find the anomaly induced
effective action \cite{2,3} (with accuracy up to conformally
invariant functional for the total effective action, see \cite{5} 
for a review).  There is no consistent approach to calculate this 
conformally invariant functional in closed form. However, one can 
find this functional as some expansion of Schwinger--DeWitt 
type \cite{6} keeping only the leading term.
Then, the effective action may be written in the following form
\cite{3,6} (these works were related with only scalars)
\bea
\label{e4}
W&=&-{1 \over 8\pi}\int d^2x \sqrt{-g}\,\left[
{N+M \over 12}R{1 \over \Delta}R
- N \nabla^\lambda \phi
\nabla_\lambda \phi {1 \over \Delta}R \right. \nn
&& \left. +\left(N + {2M \over 3} \right)\phi R
+2N\ln\mu_0^2 \nabla^\lambda \phi \nabla_\lambda \phi
\right]\ .
\eea
where $\Delta$ is two--dimensional laplacian, $\mu_0^2$ is a dimensional
parameter. 
Here, the first term represents the Polyakov anomaly 
induced action, the second and third terms give the dilaton 
dependent corrections to the anomaly induced action 
 The last 
term (conformally invariant functional) is found in ref.\cite{6}. 
Note that EA for dilaton coupled spinor has been found in ref.\cite{NNO}
 where it was shown that unlike to scalar case there is no 
ambiguety in EA related with conformally invariant 
functional. Spinor EA is known exactly in both cases:
minimal or dilaton coupled spinor.

Working in the conformal gauge
\be
\label{e5}
g_{\pm\mp}=-{1 \over 2}\e^{2\rho}\ ,\ \
g_{\pm\pm}=0,
\ee
the equations of motion may be obtained by the variation of
$\Gamma=S_{red}+W$ with respect to
$g^{\pm\pm}$, $g^{\pm\mp}$ and $\phi$
\bea
\label{e6}
0&=&{\e^{-2\phi} \over 4G}\left(
2\partial_r \rho\partial_r\phi + \left(\partial_r\phi\right)^2
-\partial_r^2\phi\right) \nn
&& -{N+M \over 12}\left( \partial_r^2 \rho - (\partial_r\rho)^2 \right)
- {N \over 2} \left(\rho+{1 \over 2}\right) (\partial_r\phi)^2 \nn
&& - {N + {2M \over 3} \over 4}
\left( 2 \partial_r \rho \partial_r \phi
- \partial_r^2 \phi \right) -{N \over 4}\ln \mu_0^2 (\partial_r\phi)^2
+ N t_0 \\
\label{e7}
0&=&{\e^{-2\phi} \over 8G}\left(2\partial_r^2 \phi
-4 (\partial_r\phi)^2 - 2\Lambda \e^{2\rho} +2 \e^{2\rho+2\phi}\right) \nn
&& +{N+M \over 12}\partial_r^2 \rho +{N \over 4}(\partial_r \phi)^2
-{N + {2M \over 3} \over 4}\partial_r^2\phi \\
\label{e8}
0&=& -{\e^{-2\phi} \over 4G}\left(-\partial_r^2\phi
+(\partial_r\phi)^2
+\partial_r^2 \rho+ \Lambda \e^{2\rho}\right) \nn
&& + {N \over 2} \partial_r(\rho \partial_r\phi)
+ {N + {2M \over 3} \over 4}\partial_r^2\rho  
+ {N \over 2}\ln \mu_0^2 \partial_r^2\phi \ .
\eea
Here, $t_0$ is a constant which is determined by the initial  conditions.
Below we are interested in the static solution that is why we replace
$\partial_\pm \rightarrow \pm{1 \over 2}\partial_r$
where $r$ is radial coordinate.

Furthermore, we change the radial coordinate $r$ by the new 
coordinate $x$
\be
\label{ss1}
x=\e^{-\phi}\ ,
\ee
which corresponds to the usual coordinate choice in the 
Schwarzschild metric:
\bea
\label{ss2}
ds^2&=&-\e^{2\rho}dt^2 + \e^{2\sigma}dx^2 + x^2d\Omega^2 \nn
\e^{\sigma}&=&-\e^{\rho + \phi}\left({d\phi \over dr}\right)^{-1} 
\eea
Then we find
\be
\label{ss2b}
\partial_r=\e^{\rho-\sigma}\partial_x\ , \quad 
\partial_r^2=\e^{2(\rho-\sigma)}\left(\partial_x^2 
+ (\partial_x\rho - \partial_x\phi)\partial_x\right)
\ee
and the equations (\ref{e6}), (\ref{e7}) and (\ref{e8}) 
can be rewritten as follows:
\bea
\label{ss3}
0&=& -{x \over 4G}\left(\partial_x\rho + \partial_x\phi\right)
-{N+M \over 12}\left( \partial_x^2 \rho 
- \partial_x\sigma\partial_x\rho \right)
- {N \over 2x^2}\left( \rho + {a \over 2}\right) \nn
&& + {N + {2M \over 3} \over 4}
\left(\partial_x\rho + \partial_x\phi\right)
+ N t_0\e^{2\sigma - 2\rho} \\
\label{ss4}
0&=&{x^2 \over 4G}\left(- {1 \over x}\left(\partial_x\rho 
- \partial_x\sigma\right) - {1 \over x^2} 
- \Lambda \e^{2\sigma} + {\e^{2\sigma} \over x^2}\right) \nn
&& +{N + M \over 12}\left(\partial_x^2 \rho + (\partial_x \rho)^2
- \partial_x\sigma\partial_x\rho \right) \nn
&& + {N + {2M \over 3} \over 4x} 
\left( \partial_x \rho - \partial_x\sigma \right)
- {M \over 6x^2} \\
\label{ss5}
0&=& -{x^2 \over 4G}\left( {1 \over x}\left(
\partial_x \rho - \partial_x\sigma\right)
+ \partial_x^2 \rho + (\partial_x\rho)^2 
- \partial_x\sigma\partial_x\rho 
+ \Lambda \e^{2\sigma}\right) \nn
&& + {N \over 4} \left\{-{2 \over x}\partial_x \rho  
+2\rho\left({1 \over x^2} - {1 \over x}
\left(\partial_x\rho - \partial_x\sigma\right)\right) 
\right\} \nn
&& + {N + {2M \over 3} \over 4} 
\left\{\partial_x^2\rho +  (\partial_r\rho)^2
- \partial_x\sigma\partial_x\rho \right\} \nn
&& + {Na \over 2}\left({1 \over x^2} 
- {1 \over x}\left(\partial_x\rho - \partial_x\sigma
\right)\right) \ .
\eea
Here $a\equiv\ln \mu_0^2 $.  Combining (\ref{ss3}) 
and (\ref{ss4}), we obtain the following equation
\bea
\label{ss7}
0&=&{x^2 \over 4G}\left( -{1 \over x^2}
-{2 \over x}\partial_x\rho - \Lambda \e^{2\sigma} 
+ {\e^{2\sigma} \over x^2}\right) \nn
&& + {N+M \over 12}(\partial_x \rho)^2 
- {N \over 2x^2} \rho + {N + {2M \over 3} \over 2x} \partial_x \rho
-{Na \over 4x^2}\e^{2\rho} + Nt_0\e^{2\sigma -2\rho} .
\eea 
This last equation is necessary to delete the constant
$t_0$ (which is defined by initial conditions) 
from further consideration. In the next sections 
we use above equations of motion in order to find
quantum corrections to different BH configurations.
Note also that there were also attempts in refs.\cite{SWH,3,6,
BRM}
 to apply such EA (usually without logarithmic term 
and only for scalars) for quantum considerations around BHs
(for semiclassical stress tensor with dilaton,see ref.\cite{LMR}). 
Notice also that some terms of above EA have similarities 
with counterterms  added by hands
 to CGHS dilatonic gravity model \cite{CGHS}, forming its 
extension as RST model \cite{RST}. For a very 
incomplete list of references on the study of 2d dilatonic 
 BHs in these models, see refs.\cite{23}.

\section{Quantum corrections to 2d and 4d Schwarzschild black hole}

First we consider the case $\Lambda=0$. In the classical limit 
($N\rightarrow 0$), we obtain, of course, the Schwarzschild 
black hole as solution of equations of motion:
\bea
\label{sb1}
&&\e^{2\rho}=\e^{2\rho_0}\ ,\quad 
\e^{2\sigma}=\e^{2\sigma_0}\ ,\nn 
&& \e^{2\rho_0}=\e^{-2\sigma_0}= 1 - {\mu \over x}\ .
\eea
Here $\mu=2GM_{\rm BH}$ and $M_{\rm BH}$ is the black hole mass. 
We now consider the quantum corrections by regarding $GN$ is 
small and assuming
\be
\label{qc1}
\rho=\rho_0 + GN\Delta_\rho\ ,\quad \sigma=\sigma_0
+ GN \Delta_\sigma\ (\sigma_0=-\rho_0 )\ .
\ee
Then  substituting (\ref{qc1}) into (\ref{ss3}), 
we obtain
\bea
\label{qc1b}
0&=&-{x \over 4}\partial_x\left(\Delta_\rho + \Delta_\sigma\right)
- {A \over 12}\left(\partial_x^2\rho_0 - \partial_x\sigma_0
\partial_x\rho_0\right) \nn
&& - \left({\rho_0 \over 2} + {a+B-1 \over 4}\right){1 \over x^2}
+ {B \over 4x}\left(\partial_x \rho_0 + \partial_x \sigma_0\right)
\nn
&& + t_0\e^{2\sigma_0 - 2 \rho_0} + {\cal O}(GN)\ .
\eea
 Substituting the classical solution (\ref{sb1}) into 
(\ref{qc1b}), we obtain
\bea
\label{sb2}
0&=&\partial_x\left(\Delta_\rho + \Delta_\sigma\right)
+ {A \over 3x}\left\{-{1 \over 4(x-\mu)^2} + {3 \over 4x^2}
- {1 \over 2x(x-\mu)}\right\} \nn
&& +\left\{\ln(x-\mu) - \ln x + a + B -1\right\}
- {4t_0 x \over (x - \mu)^2}
\eea
Here
\be
\label{sb3}
A\equiv {N+M \over N}\ , \quad 
B\equiv {N + {2M \over 3} \over N}\ .
\ee
Eq.(\ref{sb2}) can be easily integrated to give 
\bea
\label{sb4}
&& \Delta_\rho + \Delta_\sigma \nn
&& \ =A\left\{-{1 \over 12\mu(x-\mu)} + {1 \over 12\mu^2}\ln (x-\mu)
\right. \nn
&& \ \ \left. +{1 \over 8x^2} + {1 \over 6\mu x}
- {1 \over 12\mu^2}\ln {x \over l} \right\} \nn
&& \ \ + {a+B-1 \over 2x^2} - {1 \over 4x^2} - {1 \over 2\mu x}
+ {1 \over 2}\left({1 \over x^2} - {1 \over \mu^2}\right)
\ln\left( 1 - {\mu \over x}\right) \nn
&& \ \ +4t_0\left\{-{\mu \over x-\mu} + \ln (x - \mu)\right\}\ .
\eea
Here $l$ is a constant of the integration. 

On the other hand,  substituting (\ref{qc1}) into (\ref{ss7}), 
we get
\bea
\label{qc3b}
0&=&-{x \over 2}\partial_x \Delta_\rho + {1 \over 2}
\e^{2\sigma_0}\Delta_\sigma 
+ {A \over 12}\left(\partial_x\rho_0\right)^2 
- {1 \over 2x^2}\rho_0 \nn
&& + {B \over 2x}\partial_x\rho_0 - {a \over 4x^2}
+ t_0\e^{2\sigma_0 - 2\rho_0} + {\cal O}(GN)\ .
\eea
Using (\ref{qc1b}) and (\ref{qc3b}), we can delete $\Delta_\rho$ 
and obtain
\bea
\label{qc3c}
0&=&{x \over 2}\partial_x \Delta_\sigma 
+ {1 \over 2}\e^{2\sigma_0}\Delta_\sigma \nn
&& + {A \over 12}\left(2\partial_x^2\rho_0 
- 2\partial_x\sigma_0\partial_x\rho_0 
+ \left(\partial_x\rho_0\right)^2 \right) \nn
&& + \left({\rho_0 \over 2} + {a - B +1 \over 4}
\right){1 \over x^2} \nn
&& - {B \over 2x}\partial_x\sigma_0 
- t_0\e^{2\sigma_0 - 2\rho_0} + {\cal O}(GN)\ .
\eea
Substituting the classical solution (\ref{sb1}) into 
(\ref{qc3c}), one gets
\bea
\label{sb5}
0&=&\partial\left((x-\mu)\Delta_\sigma\right) 
+ A\left\{-{1 \over 24x(x-\mu)} + {1 \over 24x^2} 
- {7\mu \over 24x^3}\right\} \nn
&& +{1 \over 2}\left({1 \over x^2}-{\mu \over x^3}\right)
\left(a-B+1 + \ln (x-\mu) - \ln x\right) \nn
&& + {B\mu \over 2x^3} - {2t_0 x \over x-\mu}\ .
\eea
Integrating (\ref{sb5})
\bea
\label{sb6}
&& \Delta_\sigma = {1 \over 3(x-\mu)} \nn
&& \ \times \left\{ \Delta_S 
+ A\left( {1 \over 8\mu}\ln (x-\mu) - {1 \over 8\mu}\ln {x \over l} 
+ {1 \over 8x} - {7\mu \over 16x^2}\right) \right. \nn
&& \ + {a-B+1 \over 2}\left( {3 \over x} - {3\mu \over 2x^2} \right) 
\nn
&& \ -{3B\mu \over 4x^2} 
- {3\mu \over 4}\left({1 \over x} - {1 \over \mu}\right)^2
\ln \left( 1 - {\mu \over x}\right)  
- {3 \over 4x} + {3\mu \over 8x^2} \nn
&& \ \left. + 6t_0\left(x + \mu\ln(x-\mu)\right)\right\} \ .
\eea
Here $\Delta_S$ is a constant of the integration. 

Near the classical horizon $x\sim\mu$, $\Delta_\sigma$ 
in (\ref{sb6}) behaves as
\be
\label{sb7}
\Delta_\sigma\sim {1 \over 3(x-\mu)}
\left[\left({A \over 8\mu} + 6t_0\mu\right)\ln (x-\mu) 
+ \mbox{regular terms}\right]\ .
\ee
The singularity coming from $\ln (x-\mu)$ vanishes if we 
choose\footnote{
This procedure tells that $t_0$ here corresponds to the constants 
$C$ and $D$ of the integration in \cite{BRM}.} 
\be
\label{sb8}
t_0=-{A \over 48 \mu^2}\ .
\ee
In the choice of (\ref{sb8}), Eqs. (\ref{sb4}) and (\ref{sb6}) 
have the following forms:
\bea
\label{qs1}
&& \Delta_\rho + \Delta_\sigma \nn
&& \ =A\left\{{1 \over 8x^2} + {1 \over 6\mu x}
- {1 \over 12\mu^2}\ln {x \over l} \right\} \nn
&& \ \ + {a+B-1 \over 2x^2} - {1 \over 4x^2} - {1 \over 2\mu x}
+ {1 \over 2}\left({1 \over x^2} - {1 \over \mu^2}\right)
\ln\left( 1 - {\mu \over x}\right) \ ,\nn
\label{qs2}
&& \Delta_\sigma = {1 \over 3(x-\mu)} \nn
&& \ \times \left\{ \Delta_S 
+ A\left( -{1 \over 8\mu}\ln {x \over l}
+ {1 \over 8x} - {7\mu \over 16x^2}\right) 
+ {a-B+1 \over 2}\left( {3 \over x} - {3\mu \over 2x^2} \right)
\right. \nn
&& \left. \ -{3B\mu \over 4x^2} 
- {3\mu \over 4}\left({1 \over x} - {1 \over \mu}\right)^2
\ln \left( 1 - {\mu \over x}\right)  
- {3 \over 4x} + {3\mu \over 8x^2} + {3x \over 2\mu^2}\right\}\ .
\eea
There are  undetermined parameters $l$ and $\Delta_S$ coming from the 
constants of the integration. We now assume the radius $L$ of 
the universe is large $L\gg\mu$ but finite and we require
\be
\label{sb8b}
\Delta_\rho=\Delta_\sigma=0\quad\mbox{when}\ x=L\ .
\ee
Then we find
\be
\label{sb8c}
l=L\ ,\quad \Delta_S=-{3L \over 2\mu^2}\ .
\ee

Near the classical horizon, $\Delta_\rho+\Delta_\sigma$ is regular:
\bea
\label{sb9}
&\Delta_\rho+\Delta_\sigma & \sim B_S \nn
&B_S\equiv &{A \over \mu^2}\left({7 \over 24} 
- {1 \over 12}\ln {\mu \over l}
\right) + {a+B-1 \over 2\mu^2} - {3 \over 4\mu^2} 
\eea
and $\Delta_\sigma$ behaves as
\bea
\label{sb10}
&\Delta_\sigma&\sim {C_S \over x-\mu} \nn
&C_S\equiv &-{\Delta_S \over 3} + {A \over 24\mu}\ln{\mu \over l}
+{5A \over 48\mu} - {a-B+1 \over 4\mu} + {B \over 4\mu} 
- {3 \over 8\mu} \ .
\eea 
Eqs.(\ref{sb9}) and (\ref{sb10}) tell that the scalar curvature, 
which is given by
\bea
\label{R}
R_4&=&R + 2 \left(\nabla\phi\right)^2 + 2\e^{2\phi} \nn
R&=&-2\e^{-2\rho}\partial_r^2\rho \nn
&=&-2\e^{-2\sigma}\left(\partial_x^2 \rho + \left(\partial_x \rho 
- \partial_x \sigma \right)\partial_x \rho \right) \ ,
\eea
is regular when $x\sim \mu$.
The horizon defined by $\e^{2\rho}=0$, which corresponds 
to $x=\mu$ in the classical limit is given by 
\bea
\label{sb11}
0&=&\e^{2\rho} \nn
&\sim& \e^{2\rho_0}\left(1 + 2GN\Delta_\rho \right) \nn
&\sim& {1 \over \mu}\left(x - \mu + 2GNC_S \right) \ .
\eea
Then the entropy, which is defined by the area of horizon divided 
by $4G$, is given by
\bea
\label{sb12}
S&\sim&{\pi \over G}(x\mu - 2GNC_S )^2 \nn
&\sim&{\pi \mu^2 \over G} - 4\pi NC_S x\mu \nn
&=& {\pi\mu^2 \over G} 
- \pi N \left(-{4\mu\Delta_S \over 3} + {A \over 6}\ln{\mu \over l}
+{5A \over 12} - a + 2B - {5 \over 2} \right)\ .
\eea 
The second term in (\ref{sb12}) corresponds to the quantum 
correction.
The equation (\ref{sb11}) tells that the behavior of the metric 
near the horizon $x= \mu - 2GNC_S$ is given by 
\bea
\label{sb13}
ds^2&=& - \e^{2\rho} dt^2 
+ \e^{-2\rho +2GN(\Delta_\rho + \Delta_\sigma)}dx^2 + x^2 d\Omega^2 \nn
&\sim& - {x - \mu + 2GNC_S \over \mu} dt^2 
+ {\mu(1 + 2GNB_S) \over x - \mu + 2GNC_S} dx^2 + x^2 d\Omega^2\ .
\eea
Therefore the temperature $T$ (in the following, we put 
the Boltzmann constant $k$ to be unity, $k=1$) is given by
\bea
\label{qs8}
T&\sim&{1 \over 4\pi\mu}\left\{1 + GN\left({2C_S \over \mu}
- B_S\right)\right\} \nn
&=&{1 \over 4\pi\mu}\left[ 1 + GN\left\{-{2 \over 3\mu}\Delta_S 
+ {A \over 12\mu^2}\left(-1 + 2\ln{\mu \over l}\right) 
- {a \over \mu^2} + {B \over 2\mu^2}\right\}\right]\ .
\eea
The second term corresponds to the quantum correction.
Finally in this section, we consider the thermodynamical mass $E$, 
which is defined by
\be
\label{sb14}
dE=TdS\ .
\ee
 Using the parameter $\mu$, (\ref{sb14}) can be rewritten as
\bea
\label{sb15}
E&=&\int d\mu T{dS \over d\mu} \nn
&=&{\mu \over 2G} + N\left(-{L \over 2\mu^2} 
- {A \over 12\mu}\ln{\mu \over L} + {a \over 2\mu} 
+ {B \over 4\mu}\right)\ .
\eea
Here we used (\ref{sb8c}), (\ref{sb12}) and (\ref{qs8}). We also 
put $k=1$. The first term expresses the usual classical mass 
$M_{\rm BH}$ since $\mu=2GM_{\rm BH}$ and the second term is 
due to the quantum effects. Notice the regularization scheme 
dependence vie the presence of parameter a in above expressions. 
It maybe fixed by the choice of physical regularization. 

The qualitative structures of the entropy (\ref{sb12}), the 
temperature (\ref{qs8}) and the energy (mass) (\ref{sb15}) are 
similar to the corresponding quantities found in \cite{BRM}.
Note however that our result incorporates the quantum effects of
not only scalars but also of  
 spinors to BH configuration under consideration.

Let us turn now to 2d formulation of above results.
The classical black hole solution (\ref{sb1}) can be 
regarded as a purely two dimensional object if we start 
with the reduced action (\ref{e3}). The 
thermodynamical quantities as the energy and entropy for 
the two dimensional charged black hole with dilaton 
are evaluated on the classical 
and one-loop levels in \cite{6,MK}. In \cite{MK}, the boundary 
of the universe is introduced at the radius $r=L$ as in 
(\ref{sb8b}). If there is a boundary, we need to add the boundary 
terms to the action in order that the variation with respect to 
the metric should be well-defined. By including the boundary 
term, the formula for the energy (mass) of the black hole with
one-loop quantum correction was derived in \cite{MK} as follows,
\bea
\label{mk1}
E(\mu,L)&=&-{\e^{\lambda(L)} \over G} g^{1 \over 2}(L) 
D'(\phi(L)) \nn
&& - {\hbar \over 3}\e^{\lambda_{CL}(L)} g_{CL}^{-{1 \over 2}}(L)
g_{CL}'(L)
- {c\hbar \over 6}\e^{\lambda_{CL}(L)} g_{CL}^{1 \over 2}(L)
\phi'(L)\ .
\eea
Here $'={d \over dx}$. The quantities appeared in (\ref{mk1}) 
have the following correspondence with the quantities here,
\be
\label{mk2}
g(x)=\e^{2\rho(x)}\ ,\quad \e^{-\lambda(x)}\ , \quad
D(\phi(x))={\e^{-2\phi(x)} \over 2}={x^2 \over 2} \ .
\ee
Since only the quantum correction from the one scalar field was 
evaluated in \cite{MK}, we need to do the following replacement
\be
\label{mk3}
\hbar\rightarrow NA\ ,\ \ c\rightarrow {12B \over A}\ .
\ee
We should note that we usually choose $\hbar=1$ in this paper. 
In the expression (\ref{mk1}), the quantities with suffices $CL$ 
correspond to the classical ones. 
If we choose the boundary condition as in (\ref{sb8b}), we find 
\be
\label{mk4}
g(L)=g_{CL}(L)=1-{\mu \over L}\ ,\ \ 
\Lambda(L)=\Lambda_{CL}(L)=0\ .
\ee
Therefore 
\be
\label{mk5}
E(\mu, L)=-{L \over G}\left(1-{\mu \over L}\right)^{1 \over 2}
-{NA\mu \over 3L^2}\left(1-{\mu \over L}\right)^{-{1 \over 2}}
+{2NB \over L}\left(1-{\mu \over L}\right)^{1 \over 2}\ .
\ee
Since the energy (\ref{mk5}) diverges in the limit of 
$L\rightarrow +\infty$, we need to subtract $E(0,L)$ before 
taking the limit of $L\rightarrow +\infty$. Then we obtain
\be
\label{mk6}
E_{sub}\equiv\lim_{L\rightarrow +\infty}
\left(E(\mu,L)-E(0,L)\right)={\mu \over 2G}\ ,
\ee
which is nothing but the classical  black hole 
mass. This tells that there is no the quantum correction for 
the mass or the boundary condition in (\ref{sb8b}) corresponds 
to the renormalization condition that the mass does not suffer 
the quantum correction. 
Even if the black hole is purely two dimensional one
(but with dilaton), the 
definition of the temperature is not changed and one gets
(\ref{qs8}).
Then using the energy (\ref{mk6}), the temperature (\ref{qs8}) 
and the definition of the entropy (\ref{sb14}), we obtain the 
following expression of the entropy
\bea
\label{mk7}
S&=&\int {dE \over T}=\int {d\mu \over T(\mu)} {dE \over d\mu} \nn
&=&{\pi \mu^2 \over G}- 2\pi N\left\{-{L \over \mu}+ 
\left\{-{A \over 12} - a + {B \over 2}\right)\ln{\mu \over l}
+ {A \over 12}\left(\ln {\mu \over L}\right)^2 + c \right\} \nn
&& + {\cal O}(L^{-1}) + {\cal O}\left(GN)^2\right)\ .
\eea
Here $c$ is the constant of the integration. The classical part 
coincides with the usual Bekenstein-Hawking entropy when we regard 
the black hole as the four dimensional object.

Hence we calculated quantum corrections to simplest black hole 
thermodynamical quantities. The result is actually obtained for
two objects: 4d  BH and the same object described as 2d 
dilatonic BH.  
That shows remarkable property of s-wave EA that it could be applied 
to 4d as well as to 2d geometry (where it looks already as 
complete EA). It is straitforward now to extend 
the discussion for other types of BHs.

\section{Quantum corrections to SdS and SAdS black holes}

We now consider more general Schwarzschild-(anti-)de Sitter black 
holes. In the classical limit ($N\rightarrow 0$) in (\ref{ss3}) 
$\sim$ (\ref{ss7}) with non-vanishing cosmological constant $\Lambda$, 
we obtain the usual Schwarzschild-(anti)de Sitter as 
solution of equations of motion
\be
\label{ss6}
\e^{2\rho}=\e^{-2\sigma}=\e^{2\rho_0} 
\equiv 1 - {\mu \over x} - {\Lambda \over 3}x^2 
=-{\Lambda \over 3x}\prod_{i=1}^3 (x - x_i)\ .
\ee
Here $\mu$ is a constant of the integration corresponding to 
the black hole mass ($\mu =2GM_{\rm BH}$). 
The parameters $x_i$ ($i=1,2,3$) 
are solutions of the equation $\e^{2\rho_0}=0$. Among $x_i$'s, 
two are real and positive if 
$\Lambda>0$ and $\mu^2 < {4 \over 9\Lambda}$ 
and they correspond to black hole and cosmological horizons 
in the Schwarzschild-de Sitter black hole. On the other hand, only 
one is real if $\Lambda<0$. The explicit form of $x_i$ is given by
\bea
\label{ss6b}
&& x_i=\left(\alpha_+\right)^{1 \over 3}
+\left(\alpha_+\right)^{1 \over 3}\ ,\quad 
\omega\left(\alpha_+\right)^{1 \over 3}
+\omega^{-1}\left(\alpha_+\right)^{1 \over 3}\ ,\quad 
\omega^{-1}\left(\alpha_+\right)^{1 \over 3}
+\omega\left(\alpha_+\right)^{1 \over 3}\ ,\nn
&& \omega=\e^{2i\pi \over 3}\ ,\quad 
\alpha_\pm={1 \over 2\Lambda}\left(-3\mu\pm\sqrt{9\mu^2 
- {4 \over \Lambda}}\right)\ .
\eea
We should also note that the solutions $\{x_i,\ i=1,2,3\}$ 
satisfy the following relations:
\bea
\label{solrel}
&& X_1 \equiv \sum_{i=1}^3 x_i = 0 \nn
&& X_2 \equiv \sum_{i,j=1, i>j}^3 x_i x_j = -{3 \over \Lambda}\ ,
\quad X_3 \equiv \prod_{i=1}^3 x_i = -{3\mu \over \Lambda}\ .
\eea
If we start from the reduced action (\ref{e3}), the classical 
solution (\ref{ss6}) can be regarded to express the 
purely two dimensional 
Schwarzschild-(anti) de Sitter black holes with dilaton. 
Hence, we again obtain 2d or 4d formulation for such object.
Note that reduction of SAdS BH maybe understood 
as 2d dilatonic AdS BH where quantum effects of dilaton
were recently discussed in ref.\cite{kim}.

As in (\ref{qc1}), we consider the quantum corrections  
regarding $GN$ as small. 
Substituting the classical solution (\ref{ss6}) into (\ref{qc1b})
and integrating it, we obtain
\bea
\label{qc2}
&& \Delta_\rho + \Delta_\sigma \nn
&& = \Delta_0 
+ A\left\{{1 \over 8x^2} 
- {1 \over 12}\sum_{i=1}^3\left(  {1 \over x_i(x - x_i)}
- {1 \over x_i^2}\ln(x-x_i) \right.\right. \nn
&& \ \left.\left. + {1 \over x_i^2 }\ln x- {2 \over x_ix}\right) 
+ {1 \over 6}\sum_{i=1}^3 {1 \over x_i Y_i}\ln (x-x_i) 
\right\} \nn
&& \ + {a'+B-1 \over 2x^2} 
- {1 \over 2x^2}\left(\ln x + {1 \over 2}\right) \nn
&& \ + \sum_{i=1}^3 \left\{{1 \over 2}\left({1 \over x^2}
-{1 \over x_i^2}\right)\ln (x-x_i)
+ {1 \over 2x_i^2}\ln x - {1 \over 2x_i x}\right\} \nn
&& \ +{36t_0 \over \Lambda^2}\sum_{i=1}^3\left\{
-{1 \over x_i Y_i^2}{1 \over x - x_i} \right. \nn
&& \ \left. + {1 \over x_i^3 Y_i^3}\left(4x_i^2 
- \prod_{k=1,k\neq i}^3x_k\right)
\ln (x-x_i)\right\}\ . 
\eea
Here $\Delta_0$ is a constant of the integration and 
\be
\label{qc3}
Y_i\equiv {1 \over x_i}\prod_{i,j=1,i\neq j}^3 (x_j - x_i) \ ,
\quad a'\equiv \ln\left(-{\Lambda \over 3}\right) + a \ .
\ee
On the other hand,  substituting (\ref{ss6}) into (\ref{qc3c}), 
one gets
\bea
\label{qc4}
\Delta_\sigma &=& {1 \over \prod_{i=1}^3(x - x_i)} \nn
&& \times \left[ \Delta_1 + A \left\{ {x \over 6} - {X_2 \over 24x}
+ {7X_3 \over 48x^2} - {1 \over 24}\sum_{i=1}^3Y_i
\ln\left(1 - {x_i \over x}\right) \right\} \right. \nn
&& +{a'-B+1 \over 2}\left( x - {X_2 \over x} + {X_3 \over 2x^2}\right) 
+ B\left(x + {X_3 \over 4x^2}\right) \nn
&& - {x \over 2}\left(\ln x - 1\right) 
+ {X_2 \over 2x}\left(\ln x + 1 \right) \nn
&& - {X_3 \over 4x^2}\left(\ln x + {1 \over 2}\right) 
+ {1 \over 2}\sum_{i=1}^3 (x-x_i)\left(\ln (x-x_i) -1 \right) \nn
&& + {X_2 \over 2}\sum_{i=1}^3\left( - {1 \over x}\ln (x - x_i)
+ {1 \over x_i}\ln\left(1 - {x_i \over x}\right)\right) \nn
&& - {X_3 \over 2}\left(-{1 \over 2x^2}\ln \left( x - x_i \right)
+ {1 \over 2x_i^2}\ln\left(1 - {x_i \over x}\right)
+ {1 \over 2x_ix} \right) \nn
&& \left. - {18t_0 \over \Lambda^2}
\sum_{i=1}^3{1 \over Y_i}\ln (x-x_i) \right]\ .
\eea
Here $\Delta_1$ is a constant of the integration. 

Let $x=x_{i=I}$ corresponds to a horizon. Then  using (\ref{qc4}),  
near the horizon $x\sim x_I$, we find
\bea
\label{qq1}
\Delta_\sigma&\sim& -{1 \over x_I Y_I}{1 \over x-x_I} 
\left[ \left\{-{A \over 24}Y_I
+ {18t_0 \over \Lambda^2}{1 \over Y_I}\right\}\ln(x-x_I) \right. \nn
&& + \mbox{regular terms}\Bigr]\ .
\eea
The singularity coming from $\ln (x-x_I)$ vanishes if we choose 
the parameter $t_0$ to be
\be
\label{qq2}
t_0=-{A\Lambda^2 Y_I^2 \over 2^4\cdot 3^3}\ .
\ee
Note that we can remove the singularity corresponding to only 
one horizon since $Y_i\neq Y_j$ in general if $i\neq j$. Therefore, 
in case of Schwarzschild de-Sitter black hole, we cannot remove the
singularities corresponding to both of black hole and cosmological 
horizons.

 Using (\ref{qc2}) and (\ref{qq2}), when $x\sim x_I$, we 
find $\Delta_\rho + \Delta_\sigma$ is regular
\bea
\label{qq3}
&& \Delta_\rho + \Delta_\sigma \sim B_I \nn
B_I&\equiv & \Delta_0 + A\left[{7 \over 24x_I^2} 
- {1 \over 12}\sum_{i=1, i\neq I}^3\left\{
\left(1 - {Y_I^2 \over Y_i^2}\right)\left(  {1 \over x_i(x_I - x_i)}
- {1 \over x_i^2}\ln(x_I -x_i) \right)\right.\right. \nn
&& \left.\left. + {1 \over x_i^2 }\ln x_I- {2 \over x_ix_I} 
- {1 \over 6}\left(\sum_{j\neq i, j=1}^3 
{1 \over x_i(x_i - x_j)}\right)
\ln (x_I-x_i) 
\right\} - {1 \over 12x_I^2}\ln x_I\right] \nn
&& + {a' +B-1\over 2x_I^2} - {3 \over 4x_I^2} \\
&& + \sum_{i=1, i\neq I}^3 \left\{{1 \over 2}\left({1 \over x_I^2}
-{1 \over x_i^2}\right)\ln (x_I-x_i)
+ {1 \over 2x_i^2}\ln x_I - {1 \over 2x_i x_I}\right\} \ . 
\nonumber
\eea
Then  using (\ref{qc4}) and (\ref{qq2}), we find 
\bea
\label{qq4}
\Delta_\rho&\sim& -\Delta_\sigma \nn
&\sim& {C_I \over x - x_I} +\ \mbox{regular terms} \nn
C_I&\equiv& {1 \over \prod_{i=1,i\neq I}^3(x - x_i)} \nn
&& \times \left[ \Delta_1 + A \left\{ {x_I \over 6} 
- {X_2 \over 24x_I} + {7X_3 \over 48x_I^2} 
- {1 \over 24}\sum_{i=1, i\neq I}^3\left(Y_i - 
{Y_I^2 \over Y_i}\right)\ln\left(1 - {x_i \over x_I}
\right) \right\} \right. \nn
&& +{a'-B+1 \over 2}\left( x_I - {X_2 \over x_I} 
+ {X_3 \over 2x_I^2}\right) 
+ B\left(x_I + {X_3 \over 4x_I^2}\right) \nn
&& - {x_I \over 2}\left(\ln x_I - 1\right) 
+ {X_3 \over 4}\left(\ln x_I\right)^2 
+ {X_2 \over 2x_I}\left(\ln x_I + 1 \right) \nn
&& - {X_3 \over 4x_I^2}\left(\ln x_I + {1 \over 2}\right) 
+ {1 \over 2}\sum_{i=1,i\neq I}^3 
(x_I -x_i)\left(\ln (x_I -x_i) -1 \right) \nn
&& + {X_2 \over 2}\sum_{i=1,i\neq I}^3
\left( - {1 \over x_I}\ln (x_I - x_i)
+ {1 \over x_i}\ln\left(1 - {x_i \over x_I}\right)\right) 
- {X_2 \over 2x_I}\ln x_I \nn
&& - {X_3 \over 2}\left(-{1 \over 2x_I^2}
\ln \left( x_I - x_i \right)
+ {1 \over 2x_i^2}\ln\left(1 - {x_i \over x_I}\right)
+ {1 \over 2x_ix_I} \right) \nn
&& \left. - {X_3 \over 2}\left( -{1 \over 2x_I^2}\ln x_I 
+ {1 \over 2x_I^2}\right) \right]\ .
\eea
Eq.(\ref{qq4}) tells that the scalar curvature in (\ref{R}) 
is regular when $x\sim x_I$.
The horizon defined by $\e^{2\rho}=0$, which corresponds 
to $x=x_I$ in the classical limit is given by 
\bea
\label{qc8}
0&=&\e^{2\rho} \nn
&\sim& \e^{2\rho_0}\left(1 + 2GN\Delta_\rho \right) \nn
&\sim& - {\Lambda \over 3}\left(
\prod_{j=1,j\neq I}^3(x_j - x_I)\right)\nn
&& \times \left( x_I - 2GN C_I\right)^{-1}
\left(x - x_I + 2GN C_I\right) \ .
\eea
Then the entropy, which is defined by the area of horizon divided 
by 4, is given by
\bea
\label{qc10}
S&\sim&\pi(x_I - 2GNC_I )^2 \nn
&\sim&\pi x_I^2 - 4\pi GNC_I x_I \ .
\eea 
The second term in (\ref{qc10}) corresponds to the quantum 
correction.
When we regard the black holes as purely two dimensional ones  
starting from (\ref{e3}), it is rather difficult to define the entropy 
since horizon is a point and does not have an area. In case of 
Schwarzschild de Sitter black holes, it is also difficult to 
define the entropy in the same way 
as in Schwarzschild case in (\ref{mk7}). 
The difficulty comes since we cannot define the black hole mass 
as in (\ref{mk6}), where we need to choose the radius $L$ of the 
universe large, which non-trivial cosmological horizon prevents. 

The equation (\ref{qc8}) tells that the behavior of the metric 
near the horizon $x= x_I - 2GNC_I$ is given by 
\bea
\label{qc11}
ds^2&=& - \e^{2\rho} dt^2 
+ \e^{-2\rho +2GN(\Delta_\rho + \Delta_\sigma)}dx^2 + x^2 d\Omega^2 \\
\e^{2\rho}&\sim& - {\Lambda \over 3}\left(\prod_{j=1, j\neq I}^3 
\left(x_j - x_I + 2GN C_I \right)\right) \nn
&& \times \left( x_I - 2GN C_I\right)^{-1}
\left(x - x_I + 2GN C_I\right) \nn
&\sim&- {\Lambda \over 3}\left\{Y_I
+\left(- 6 + {2Y_I \over x_I}\right) GN C_I \right\}\nn
&& \times \left(x - x_I + 2GN C_I\right) \ .\nonumber
\eea
Therefore the temperature $T$ is given by
\be
\label{qc12}
T\sim \left| {\Lambda \over 12\pi}\left\{Y_I
+\left(- 6 + {2Y_I \over x_I}\right)GN C_I
-GNY_IB_I \right\}\right| \ .
\ee
The second and third terms correspond to the quantum correction.
The definition of the temperature is the same if we regard the black 
holes as purely two dimensional ones. We should note that there are 
two different temperatures corresponding to the black hole and the 
cosmological horizons, respectively, for Schwarzschild de Sitter 
black holes.

The Schwarzschild black hole in the previous section can be 
obtained as a limit of $\Lambda\rightarrow +0$ in 
Schwarzschild-de Sitter black hole.
In the limit, $x_i$'s in (\ref{ss6}) behave as 
\be
\label{qs0}
x_i\rightarrow\ \mu, \pm \sqrt{3 \over \Lambda}\ .
\ee
Here the constant of the integration $l$ is related with $\Delta_0$ 
by $\Delta_0= {1 \over 12\mu^2}\ln l$ and $\Delta_S$ is given by 
$\Delta_S= \Lambda\Delta_1+ \mbox{const.}$ The additional 
constant diverges in the limiting procedure $\Lambda\rightarrow +0$ 
but the divergence can be absorbed into the redefinition of 
$\Delta_S$. 

We now consider de Sitter space as a limit of $\mu\rightarrow 0$.
If we choose $x_I$ as a cosmological horizon
\be
\label{ds1}
x_I=h\equiv \sqrt{3 \over \Lambda}\ ,
\ee
$t_0$ in (\ref{qq2}) has the following form:
\be
\label{sd2}
t_0=-{A\Lambda \over 36}=-{A \over 12h^2}\ .
\ee
Then we obtain
\bea
\label{sd3}
&& \Delta_\rho + \Delta_\sigma \nn
&& \ = \tilde\Delta_0 - {A \over 6x^2} + {a+B-1 \over 2x^2} \nn
&& \ \ + {1 \over 2}\left({1 \over x^2} - {1 \over h^2}\right)
\ln \left( 1 - {x^2 \over h^2} \right) 
+ {1 \over h^2}\ln {x \over h} \\
\label{sd4}
&& \Delta_\sigma \nn
&& \ = {1 \over x\left(x^2 - h^2\right)}\left\{
\tilde\Delta_1 + {Ax \over 6} 
+ {a-B+1 \over 2}\left(x + {h^2 \over x}\right) \right. \nn
&& \left. +Bx + {(x-h)^2 \over 2x}\ln\left(1-{x \over h}\right)
+ {(x+h)^2 \over 2x}\ln\left(1+{x \over h}\right)\right\}\ .
\eea
Here the constants $\tilde\Delta_0$ and $\tilde\Delta_1$ 
are related with the constants of the integration 
$\Delta_0$ and $\Delta_1$ by
\be
\label{sd5}
\tilde\Delta_0=\Delta_0 + {1 \over 2h^2}\ln (-1)\ ,\quad
\tilde\Delta_1=\Delta_1 + h\ .
\ee
The expressions (\ref{sd3}) and (\ref{sd4}) can be used for 
anti-de Sitter space by analytically continuing $h$ by 
$h\rightarrow i\sqrt{-{3 \over \Lambda}}$:
\bea
\label{a4}
&& \Delta_\rho + \Delta_\sigma \nn
&& \ = \tilde\Delta_0 - {A \over 6x^2} + {a+B-1 \over 2x^2} \nn
&& \ \ + {1 \over 2}\left({1 \over x^2} + {1 \over h^2}\right)
\ln \left( 1 + {x^2 \over h^2} \right) 
- {1 \over h^2}\ln {x \over h} \\
\label{a5}
&& \Delta_\sigma \nn
&& \ = {1 \over x\left(x^2 + h^2\right)}\left\{
\tilde\Delta_1 + {Ax \over 6} 
+ {a-B+1 \over 2}\left(x - {h^2 \over x}\right) \right. \nn
&& \left. +Bx + {x^2 -h^2 \over 2x}\ln\left(1+{x^2 \over h^2}\right)
+ 2h{\rm arctan}\left({x \over h}\right)\right\}\ .
\eea
We should also note that there appears a singularity at the origin 
$x=0$, which might be a sign that the $s$-wave approximation is not 
valid there since the origin is always a singularity of $s$-wave. 
The $s$-wave approximation, however, should be valid near the 
cosmological horizon.

Near the classical horizon $x\sim h$, we find 
\bea
\label{sd6}
\Delta_\rho + \Delta_\sigma &\sim& \tilde B \nn
\tilde B &\equiv& \tilde\Delta_0 - {A \over 6h^2} 
+ {a + B - 1 \over 2h^2} \\
\label{sd7}
\Delta_\sigma &\sim& {\tilde C \over x-h} \nn
\tilde C &\equiv& {\tilde\Delta_1 + {Ah \over 6} + (a+1+2\ln 2)h 
\over 2h^2}\ .
\eea
Then  putting $x_I=h$ and $C_I=\tilde C$ in (\ref{qc10}), we find 
the expression  for the entropy $S$:
\bea
\label{sd8}
S&=&\pi h^2 - 4\pi GN \tilde C h \nn
&=&{3\pi \over \Lambda}-2\pi GN 
\left\{\tilde\Delta_1\sqrt{\Lambda \over 3} + {A \over 6} 
+ a + 1 + 2\ln 2 \right\}\ .
\eea
It is interesting to note that last expression 
describes quantum corrections to the entropy
of expanding inflationary Universe (as de Sitter 
space may be considered as such inflationary Universe).
 That gives new terms proportional 
to particles number as compare with classical 
entropy of expanding Universe discussed 
extensively in refs.\cite{Gary}.

 Taking $Y_I=2h$ and $B_I=\tilde B$ in (\ref{qc12}), 
the temperature is expressed as
\bea
\label{sd8b}
T&=&{\Lambda \over 12\pi}\left\{2h + GN\left(2\tilde C 
- 2h \tilde B\right) \right\} \\
&=&{1 \over 2\pi}\sqrt{\Lambda \over 3} + {GN\Lambda \over 12\pi}
\left\{{\Lambda\tilde\Delta_1 \over 3} 
- 2\tilde\Delta_0\sqrt{3 \over \Lambda}
+ \left({A \over 2} + 2 + 2\ln 2 - B\right)\sqrt{\Lambda \over 3}
\right\}\ .\nonumber
\eea
Hence we found quantum corrections to the temperature and the entropy 
 for 4d  de Sitter space (as the limit of SdS BH) as well as 
for SdS BH and SAdS BH. In the last case of SdS or SAdS BHs 
we also defined the quantum correction to the temperature of
corresponding 2d object (i.e. 
corresponding BH with dilaton). The calculation
of 2d quantum entropy is more difficult and cannot be done by using
only 4d point of view unlike the case of Schwarzschild  BH.

\section{Quantum corrections to 't Hooft BH model}

We now consider 't Hooft's BH  model \cite{thft}. In the model, 
the Hawking massless 
particles emitted from a black hole are treated as an 
envelope of matter which obeys the classical 
  equation of state, and acts as a 
source of the gravity. Then, the horizon vanishes and 
the black hole entropy can be calculated as that of the particles 
by using the Hartree-Fock approximation.  In this model 
BH entropy calculated with such recipe does not give
the standard 1/4 coefficient in area law.
Below we only consider scalars for simplicity, then
one can put $A=B=1$ in formulas of previous section.

In the Rindler space limit, where the black hole mass 
$M_{\rm BH} = {\mu \over 2G}$ is large, 
the metric has the following form\cite{thft}
\be
\label{thf1}
\e^{2\rho_0}={\lambda^2 \over P}
\left({2M_{\rm BH} \over x}\right)^6\ ,\ \ 
\e^{2\sigma_0}={\lambda^2 \over P^3}
\left({2M_{\rm BH} \over x}\right)^{14}\ ,
\ee
where
\be
\label{thf2}
P\equiv {1 \over 5}\left\{
\left({2M_{\rm BH} \over x}\right)^5 - 1 \right\}
\ ,\ \ \lambda\equiv {1 \over 48M_{\rm BH}}\sqrt{N \over 5\pi}\ .
\ee
In the following, we put $G=1$ when there is no any confusion.
 Following \cite{thft}, a new coordinate $y$ is introduced by
\be
\label{thf3}
y\equiv \left({2M_{\rm BH} \over x}\right)^5\ .
\ee

We suppose that matter 
quantum effects  are described 
by the same effective action as in section 2 (without spinors).
Then one can use the same effective action as quantum correction 
to 't Hooft equations of motion. That, of course,
adds new terms to these equations.

To be more specific, 
 substituting the classical solution (\ref{thf1}) into 
(\ref{qc1b}) and subtracting the contribution from 
the classical matter, 
we obtain
\bea
\label{thf4}
&& \Delta_\rho + \Delta_\sigma \nn
&& = {1 \over (2M_{\rm BH})^2}\left[ \Delta_0' -2 y^{2 \over 5} 
+ {3 \over 5}F_{2 \over 5}(y) 
+ {5 \over 12}{y^{7 \over 5} \over y-1} \right. \nn
&& \ +{3 \over 5}y^{2 \over 5}\left(\ln y - {5 \over 2}\right) 
- {1 \over 2}y^{2 \over 5}\ln (y-1) + {a_1 \over 2} y^{2 \over 5} \nn
&& \ \left. -80 M_{\rm BH}^2 t_0\left(-{y^{3 \over 5} \over y-1} 
+ {3 \over 5}F_{-{2 \over 5}}(y)\right)\right]\ .
\eea
Here $\Delta_0'$ is a constant of the integration and a constant 
$a_1$ is defined by
\be
a_1\equiv a + \ln \left(5\lambda^2\right)\ .
\ee
We also introduced a function $F_\alpha (y)$ which is defined by
\be
\label{thf5}
F_\alpha(y)\equiv 
\sum_{n=0}^\infty {1 \over \alpha - n}y^{\alpha -n}\ ,
\ee
which appears in the integration of 
$\int^y dy {y^\alpha \over y-1}$. The constants of the 
integration are always absorbed into the definition of 
$\Delta_0'$ in (\ref{thf4}) (and $\Delta_1'$ in the later 
eq.(\ref{thf9b})). 

When $y\sim 1$, we find the behavior of 
$\Delta_\rho + \Delta_\sigma$ from (\ref{thf5})
\be
\label{thf6}
\Delta_\rho + \Delta_\sigma \sim \left({5A \over 12} 
+ 80 M_{\rm BH}^2 t_0\right){ 1 \over y-1}\ .
\ee
 Requiring that the singularity vanishes, we can determine 
the constant $t_0$ by
\be
\label{thf7}
t_0=-{1 \over 2^6\cdot 3M_{\rm BH}^2}\ .
\ee
On the other hand,  substituting the classical solution 
(\ref{thf1}) into (\ref{qc3c}) (and subtracting the contribution 
from the classical matter again), we obtain
\bea
\label{thf8}
\Delta_\sigma&=&
{\e^{-G(y)} \over (2M_{\rm BH})^2}\int^y dy \, \e^{G(y)}
\left[-{23 \over 10}y^{-{3 \over 5}} 
+ {8 \over 3}{y^{2 \over 5} \over y-1} 
- {5 \over 24}{y^{7 \over 5} \over (y-1)^2}  \right.\nn
&& + \left({3 \over 25}\ln y - {1 \over 10}\ln (y-1) + 
{a_2 \over 10}\right)y^{-{3 \over 5}} 
\left. - 40 M_{\rm BH}^2 t_0 {y^{3 \over 5} \over (y-1)^2}\right]\ .
\eea
Here
\bea
\label{thf9}
G(y)&\equiv& -25\lambda^2 \left\{ - {y^{9 \over 5} \over 2(y-1)^2}
- {9 \over 10}{y^{4 \over 5} \over y-1} 
+ {18 \over 25}F_{-{1 \over 5}}(y)\right\} \nn
&=&-25\lambda^2 \left[ - {y^{9 \over 5} \over 2(y-1)^2}
- {9 \over 10}{y^{4 \over 5} \over y-1} \right. \nn
&& + {18 \over 25}\left\{
\sqrt{5+\sqrt{5} \over 2}\tan^{-1}\left(2\sqrt{2 \over 5+\sqrt 5}
\left({1 - \sqrt{5} \over 4} + y^{1 \over 5}\right)\right) \right. \nn
&& + \sqrt{5-\sqrt{5} \over 2}\tan^{-1}\left(2\sqrt{2 \over 5-\sqrt 5}
\left({1 + \sqrt{5} \over 4} + y^{1 \over 5}\right)\right) 
+ \ln \left(-1 + y^{1 \over 5}\right)\nn
&& - {1-\sqrt{5} \over 4}\ln\left(1+{1-\sqrt{5} \over 2}y^{1 \over 5}
+ y^{2 \over 5}\right) \nn
&& \left.\left.
- {1+\sqrt{5} \over 4}\ln\left(1+{1+\sqrt{5} \over 2}y^{1 \over 5}
+ y^{2 \over 5}\right) \right\} \right] \ .
\eea
In the limit of $M_{\rm BH}\rightarrow\infty$, $G(y)$ vanishes and 
we obtain
\bea
\label{thf9b}
\Delta_\sigma&=&{1 \over (2M_{\rm BH})^2}\left\{\Delta_1' 
-{13 \over 2}y^{2 \over 5}+{5 \over 24}{y^{7 \over 5} \over 
y-1} + {21 \over 8}F_{2 \over 5}(y)
+ {3 \over 10}y^{2 \over 5}\ln y \right.\\
&& \left. - {1 \over 4}y^{2 \over 5}\ln (y-1) + {1 \over 4}F_{2 \over 5}(y) 
+ {a_2 \over 4}y^{2 \over 5} 
- 40 M_{\rm BH}^2 t_0 \left(-{y^{3 \over 5} \over y-1} 
+ {3 \over 5}F_{-{2 \over 5}}(y)\right)\right\}\ .\nonumber
\eea
Here $\Delta_1'$ is a constant of the integration.
If we choose $t_0$ by (\ref{thf7}), Eqs.(\ref{thf4}) and 
(\ref{thf9b}) tell 
\be
\label{thf10}
\Delta_\rho\ ,\ \ \Delta_\sigma = {\cal O}\left(\ln (y-1)\right)
\quad \mbox{when} \ y\rightarrow 1+0\ .
\ee
We also find 
\be
\label{thf11}
\Delta_\rho\ ,\ \ \Delta_\sigma = {\cal O}\left(y^{2 \over 5}\right)
\quad \mbox{when} \ y\rightarrow +\infty\ .
\ee
 Using (\ref{thf4}) and (\ref{thf8}), we can estimate the entropy 
with the help of 
 the expression in \cite{thft} for massless scalar fields :
\be
\label{thf12}
S={2\pi^2 \Sigma \over 45 \beta^3}\int dy 
{H(y) \sqrt{F(y)} \over A(y)^{3 \over 2}}\ .
\ee
Here $\beta={1 \over T}$ and the metric is assumed to have 
the following form
\be
\label{thf13}
ds^2= - A(y)dt^2 + F(y)dy^2 + (2M_{\rm BH})^2 H(y)d\Omega^2
\ee
and $\Sigma$ is the area of the surface given by $y=1$, i.e., 
$\Sigma=4\pi (2M_{\rm BH})^2$. 
Eq.(\ref{thf12}) expresses the contribution from one kind of 
massless scalar field. Then total entropy is given by multiplying 
$N$ with $S$ in (\ref{thf12}). 
We now find the following expression including the quantum 
correction 
\be
\label{thf14}
S={2\pi^2\Sigma \over 45\beta^3}{2M_{\rm BH} \over 5\lambda^2}
\int_1^\infty {dy \over y^2}\left(1 + N (\Delta_\sigma 
- 3\Delta_\rho)\right)\ .
\ee
Substituting (\ref{thf4}) and (\ref{thf9b}) into (\ref{thf14}), 
we obtain
\bea
\label{thf15}
S&=&{2\pi^2 \Sigma \over 45\beta^3}{2M_{\rm BH} \over 5\lambda^2}
\int_1^\infty {dy \over y^2}\left[1 + {N \over (2M_{\rm BH})^2}\left\{
-3\Delta_0' + 4\Delta_1' \right.\right. \nn
&& + 2F_{1 \over 2}(y) - {5 \over 12}{y^{7 \over 5} \over y-1}
- {3 \over 5}y^{2 \over 5}\ln y - {31 \over 2} y^{2 \over 5} \nn
&& + {1 \over 2}y^{2 \over 5}\ln (y-1) 
+ {43 \over 10}F_{2 \over 5}(y) 
+\left( -{3 \over 2}a_1 + a_2\right)y^{2 \over 5} \\ 
&& \left.\left. 
-80M_{\rm BH}^2 t_0 \left({y^{3 \over 5} \over y-1} 
- {3 \over 5}F_{-{2 \over 5}}(y)\right) \right\}\right]\ .
\nonumber
\eea
Using (\ref{thf7}) and the following numerical results:
\bea
\label{thf16}
&& \int_1^\infty {F_{2 \over 5}(y) \over y^2}dy = 
-\int_1^\infty {F_{-{2 \over 5}}(y) \over y^2}dy = {5 \over 2}
=2.5\cdots \nn
&& \int_1^\infty {y^{-{3 \over 5}} - y^{-{7 \over 5}} \over 
y-1} dy = 1.47923\cdots \nn
&& \int_1^\infty y^{-{8 \over 5}}\ln (y-1) dy = 1.60567\cdots\ ,
\eea
we get
\bea
\label{thf16b}
S&=&{2\pi^2\Sigma \over 45\beta^3}{2M_{\rm BH} \over 5\lambda^2}
\left[1 + {N \over (2M_{\rm BH})^2}\left\{
-3\Delta_0' + 4\Delta_1' - 10.93851 \right.\right. \nn
&& \left.\left. + 1.666667\left( -{3 \over 2}a_1 + a_2\right) 
\right\}\right]\ .
\eea
In \cite{thft}, the entropy comes from the massless scalar 
fields in the background including the backreaction by  
Hawking radiation. The obtained entropy is different from 
the Bekenstein-Hawking entropy by a factor ${8 \over 5}$ 
at the classical level. Note that there are another examples
of BHs (like Kerr-bolt-AdS) where there are deviations from 
Bekenstein-Hawking law.
In (\ref{thf16b}), the correction to the entropy  
 is estimated. 
However, as quantum corrections are assumed to be small 
perturbation it was hard from the very beginning 
to expect that they may change the qualitative structure of
the model. In particular, BH entropy does not reproduce the 
Bekenstein-Hawking law.
   
If we assume ${1 \over T}=\beta=8\pi M_{\rm BH}$, 
the thermodynamical mass defined by (\ref{sb14}) can be given by
\bea
\label{thf17}
E&=&\int dT T{dS \over dT} \nn
&=&{2 \over 5N}{1 \over 4\pi T}\left[2 + 
{(4\pi)^2 N \over 2}T^3\left\{
-3\Delta_0' + 4\Delta_1' - 10.93851 \right.\right. \nn
&& \left. + 1.666667\left( -{3 \over 2}a_1 + a_2\right) 
\right\} \nn
&=&{4M_{\rm BH} \over 5N}\left[2 + 
{N \over 64 M_{\rm BH}^3}\left\{
-3\Delta_0' + 4\Delta_1' - 10.93851 \right.\right. \nn
&& \left. + 1.666667\left( -{3 \over 2}a_1 + a_2\right) 
\right\} \ .
\eea
The ``mass'' $M_{\rm BH}$ is originally the mass of the black 
hole measured in the asymptotically flat region, i.e.,  outside 
of the black hole. In this section, we can regard $M_{\rm BH}$ is 
defined by the temperature.

Hence, new methods (not semiclassical ones)
are necessary in strong quantum gravity regime in
order to solve the problems related with 't Hooft BH model.
It could be that BH-thermodynamical laws in such regime 
should be completely revised.

\section{Discussion}

We studied quantum properties of S(A)dS BHs using anomaly 
induced EA for dilaton coupled matter. The explicit evaluation 
of thermodynamic quantities (temperature, mass, entropy) with
account of quantum effects is done for S(A)dS BHs as well
as for their limiting cases: Schwarzschild and de Sitter spaces. 
The case of 't Hooft BH model is also considered.

The anomaly induced EA under discussion gives the possibility 
for 4d formulation ($s$-wave approximation) and for 
2d formulation. Hence, the corresponding results 
(with small modifications) are valid for the same background 
being quantum corrected 4d BH or quantum corrected 2d dilatonic 
BH. The formulation is general enough. There is no problem 
to apply the same technique for the calculation 
of quantum corrections to any other gravitational background.
Using of other versions of EA (say, the one induced by 4d anomaly)
does not make big qualitative difference.

Note, however, that we considered quantum EA as small 
perturbation to classical one. In this way, we estimated 
quantum corrections to the entropy of de Sitter (inflationary) 
Universe. Clearly, such investigation should be more important 
when quantum effects play the dominant role (say, gravitational 
background is induced by quantum effects). As an example, 
let us consider ${\cal N}=4$ $SU(N)$ super YM theory.
 In this case, starting  
with zero cosmological constant we get the effective 
cosmological constant \cite{brevik}: 
$H^2=-{1 \over \kappa b'}$, $b'=-{(N^2-1) \over 
(8\pi)^2}$ and $\kappa$ is gravitational constant. 
Hence, classical entropy of de Sitter Universe is  
quantum entropy on the same time and it is given by the
 standard expression with the change of cosmological 
constant by the effective cosmological constant $H^2$.
It would be really interesting to investigate the questions 
related with the quantum entropy of BHs and expanding Universe
 (when there is cosmological horizon \cite{Gary}) in such
strong quantum regime.

\ 

\noindent
{\bf Acknowledgments}. SDO would like to thank G. Gibbons, 
E. Mottola  and G.'t Hooft for useful discussion of some 
related questions. The research by
SDO was partially supported by a RFBR Grant N\,99-02-16617,
by Saxonian Min. of Science and Arts and by Graduate College
``Quantum Field Theory" at Leipzig University.


\begin{thebibliography}{99}
\bibitem{BCH} J.M. Bardeen, B. Carter and S.W. Hawking,
{\sl Comm.Math.Phys.} {\bf 31} (1973) 161.
\bibitem{BH} J.D. Bekenstein, {\sl Nuovo Cim.Lett.} {\bf 4} (1972) 737;
S.W. Hawking, {\sl Comm.Math.Phys.} (1975) 199.
\bibitem{SV} A. Strominger and C. Vafa, {\sl Phys.Lett.} {\bf B379} (1996) 99;
for a review, see A. Peet, {\sl Class.Quant.Grav.}
 {\bf 15} (1998) 3291.
\bibitem{SC} S.Carlip, hep-th/9806026.
\bibitem{BZ} A. Bytsenko, L. Vanzo and S. Zerbini,
 {\sl Phys.Rev.} {\bf D57} (1998) 4917.
\bibitem{TJ} T. Jacobson, gr-qc/9404039.
\bibitem{1} E. Elizalde, S. Naftulin and S.D. Odintsov,
{\sl Phys.Rev.} {\bf D49} (1994) 2852.
\bibitem{2} R. Bousso and S. Hawking,
{\sl Phys.Rev.} {\bf D56} (1997) 7788.
\bibitem{3} S. Nojiri and S.D. Odintsov,
{\sl Mod.Phys.Lett.} {\bf A12} (1997) 2083;
{\sl Phys.Rev.} {\bf D57} (1998) 4847.
\bibitem{4} S. Nojiri and S.D. Odintsov,
{\sl Phys.Rev.} {\bf D57} (1998) 2363;
T. Chiba and M. Siino, {\sl Mod.Phys.Lett.} {\bf A12} (1997) 709;
S. Ichinose, {\sl Phys.Rev.} {\bf D57} (1998) 6224;
A. Mikovic and V. Radovanovic,
 {\sl Class.Quant.Grav.} {\bf 15} (1998) 827;
W. Kummer, H.Liebl and D.V. Vassilevich,
{\sl Mod.Phys.Lett.} {\bf A12} (1997) 2683;
J.S. Dowker,
{\sl Class.Quant.Grav.} {\bf 15} (1998) 1881;
S. Ichinose and S.D. Odintsov, 
{\sl Nucl.Phys.} {\bf B539} (1999) 643.
\bibitem{5}
I.L. Buchbinder, S.D. Odintsov and I.L. Shapiro, Effective Action in
Quantum Gravity, IOP Publishing, Bristol and Philadelphia, 1992.
\bibitem{6}
S. Nojiri and S.D. Odintsov,
hep-th/9806055,
{\sl Phys.Rev.} {\bf D59} (1999) 044003; 
\bibitem{SWH}
R. Bousso and S.W. Hawking,
{\sl Phys.Rev.} {\bf D57} (1998) 2436;
R. Bousso,
 {\sl Phys.Rev.} {\bf D58} (1998) 083511.
\bibitem{BRM} M. Buri\'c, V. Radovanovi\'c and A. Mikovi\'c, 
gr-qc/9804083. 
\bibitem{MK} A.J.M. Medved and G. Kunstatter, 
hep-th/9904070.
\bibitem{LMR}
F. Lombardo, F.D. Mazzitelli and J. Russo,
{\sl Phys. Rev.} {\bf D59} (1999) 084002;
R. Balbinot and A. Fabbri,
 {\sl Phys.Rev.} {\bf D59} (1999) 044031;
\bibitem{NNO}
P. van Nieuwenhuizen, S. Nojiri and S.D. Odintsov,
hep-th/9901119.
\bibitem{CGHS}
C. Callan, S. Giddings, J. Harvey and A. Strominger,
{\sl Phys.Rev.} {\bf D45} (1992) 1005.
\bibitem{RST}
J. Russo, L. Susskind and L. Thorlacius,
 {\sl Phys.Lett.} {\bf B292} (1992) 13.
\bibitem{23} S.P. de Alwis, {\it Phys.Lett.} {\bf B289} (1992) 278;
A. Bilal and C. Callan, {\it Nucl.Phys.} {\bf B394} (1993) 73;
S. Nojiri and I. Oda, {\it Phys.Lett.} {\bf B294} (1992) 317;
{\it Nucl.Phys.} {\bf B406} (1993) 499;
T. Banks, A. Dabholkar, M. Douglas and M. O`Loughlin, 
{\it Phys.Rev.} {\bf D45} (1992) 3607;
R.B. Mann, {\it Phys.Rev.} {\bf D47} (1993) 4438;
D. Louis-Martinez and G. Kunstatter, {\it Phys.Rev.} {\bf D49} 
(1994) 5227;
J. Polchinski and A. Strominger, {\it Phys.Rev.} 
{\bf D50} (1994) 7403; 
T.Klobsch and T.Strobl, {\it Class.Quant.Grav.} {\bf 13} (1996) 965;
S. Bose, L. Parker and Y. Peleg, 
{\it Phys.Rev.} {\bf D52} (1995) 3512; for a review,see
A. Strominger, Les Houches lectures on black holes, 
hep-th/9501071.
\bibitem{Gary}
G. Gibbons,
 {\sl Nucl.Phys.} {\bf B292} (1987) 784 ;
{\bf B310} (1988) 636; {\bf B472} (1996) 683. 
\bibitem{thft} G. 't Hooft, {\sl Nucl.Phys.Proc.Suppl.}
 {\bf 68} (1998) 174, gr-qc/9706058 and gr-qc/9711053.
\bibitem{BSWH} 
R. Bousso and S.W. Hawking, {\sl Phys.Rev.} {\bf D54}
 (1996) 6312.
\bibitem{byts}
A. Bytsenko, S. Nojiri and S.D. Odintsov,
 {\sl Phys.Lett.} {\bf B443} (1998) 121.
\bibitem{Noj}
S. Nojiri and S.D. Odintsov, {\sl Phys.Rev.} {\bf D59}
 (1999) 044026; 
E. Elizalde, S. Nojiri and S.D. Odintsov, {\sl Phys.Rev.}
 {\bf D59} (1999) 061501.
\bibitem{brevik} I. Brevik and S.D. Odintsov,
hep-th/9902184, PLB ,to appear.
\bibitem{kim} W.T. Kim and J. Oh, hep-th/9905007.
\end{thebibliography}
\end{document}